\documentstyle[12pt]{article}

\makeatletter
\ifcase\@ptsize
  \font\tenmsy=msbm10
  \font\sevenmsy=msbm7
  \font\fivemsy=msbm5
\or
  \font\tenmsy=msbm10 scaled \magstephalf
  \font\sevenmsy=msbm8
  \font\fivemsy=msbm6
\or
  \font\tenmsy=msbm10 scaled \magstep1
  \font\sevenmsy=msbm8
  \font\fivemsy=msbm6
\fi
\newfam\msyfam
\textfont\msyfam=\tenmsy  
\scriptfont\msyfam=\sevenmsy
\scriptscriptfont\msyfam=\fivemsy
\def\Bbb{\ifmmode\let\next\Bbb@\else
\def\next{\errmessage{Use \string\Bbb\space only in math mode}}\fi\next}
\def\Bbb@#1{{\Bbb@@{#1}}}
\def\Bbb@@#1{\fam\msyfam#1}
\newfam\euffam
\font\sixeuf=eufm6
\font\eighteuf=eufm8
\font\twelveeuf=eufm10 scaled\magstep1
\textfont\euffam=\twelveeuf
\scriptfont\euffam=\eighteuf
\scriptscriptfont\euffam=\sixeuf

\makeatother

\newcommand{\BZ}{{\Bbb{Z}}}
\newcommand{\Bid}{1\!{\rm l}}

\def\pn{\par\noindent}

\def\w{{\cal W}}

\def\be{\begin{equation}}
\def\ee{\end{equation}}
\def\ba{\begin{array}}
\def\ea{\end{array}}
\def\bea{\begin{eqnarray}}
\def\eea{\end{eqnarray}}
\def\bean{\begin{eqnarray*}}
\def\eean{\end{eqnarray*}}
\def\bl{\begin{list}{}{}}

\newcommand{\reseteqn}{\setcounter{equation}{0}}
\newcommand{\mysection}{\reseteqn\section}

\if@twoside
\else
\fi
\begin{document}
  \pagestyle{empty}
  \begin{raggedleft}
IASSNS-HEP-96/69\\
hep-th/9606130\\
June 1996\\
  \end{raggedleft}
  $\phantom{x}$ 
  {\LARGE\bf
  \begin{center}
2-Dimensional Turbulence:\\ 
yet another Conformal Field Theory Solution 
  \end{center}
  }\par
  \vfill
  \begin{center}
$\phantom{X}$\\
{Mic$\hbar$ael A.I.~Flohr\footnote[1]{ email: {\tt flohr@sns.ias.edu}}}\\
$\phantom{X}$\\
{\em School of Natural Sciences\\
Institute for Advanced Study\\
Olden Lane\\
Princeton, NJ 08540, USA}
  \end{center}\par
  \vfill
  \begin{abstract}
  \noindent 
  A new conformal field theory description of two-dimensional turbulence
  is proposed. The recently established class of rational logarithmic
  conformal field theories provides a unique candidate solution which
  resolves many of the drawbacks of former approaches via minimal models.
  This new model automatically includes magneto-hydrodynamic turbulence
  and the Alf'ven effect. 
  \end{abstract}
  \vfill\vfill\vfill
  \newpage
%
%
  \setcounter{page}{1}
  \pagestyle{plain}
  \mysection{Setting the Stage for Conformal Turbulence}
  \pn
Once upon a time\footnote[1]{to be precise, September 1992} A.~Polyakov 
proposed 
a novel way of treating two-dimensional fluid mechanics: the correlation
functions of certain conformal field theories (CFTs) satisfy the Hopf
chains arising from the Navier Stokes equations \cite{Pol92}. His story has
been retold in numerous variations 
\cite{Bra95,CMU93,CST96,CoTh95,FaHa92,FeYa92,Low92,Mat92,Mor95,RaRo95,RRD95} 
and it soon became clear that
there are actually infinitely many stories of minimal models solving
two-dimensional turbulence. But all these could not lift all of the mysteries.
Moreover, different stories had different happy ends, contradicting each
other in their answers to lasting secrets as, for instance, the spectrum of 
turbulence.
  \par
What most the stories told so far have in common is that the solution to
conformal turbulence was sought among the minimal models \cite{BPZ83},
that the infrared problem was not really solved, and -- last but not least --
that the solution did not conserve any member of the infinite set of 
integrals of motion, except the very first ones. Though, to be honest, some
of them did at least notice the gaps.
  \par
I will try to tell another fairy-tale, one in which the hero is not 
a minimal model but one of the recently established rational logarithmic
conformal field theories. I will try to convince you that my hero might
have a better chance to survive the adventures settled by the above 
mentioned problems. But let us now go into medias res $\ldots$
  \par
The central point of the CFT approach to turbulence in two dimensions 
\cite{Pol92} is to consider the Hopf equations for the correlation functions 
of the velocity field $v_{\alpha}(x)$,
\be
  \sum_{p=1}^n\left\langle v_{\alpha_1}(x_1)\ldots
  \dot v_{\alpha_p}(x_p)\ldots v_{\alpha_n}(x_n)\right\rangle = 0\,,
\ee
where $\dot v_{\alpha}=\partial v_{\alpha}/\partial t$ is expressed in terms 
of $v_{\alpha}$ via the equations of motion.
It is convenient to introduce the vorticity $\omega$ and the stream function
$\psi$ given by
\be
  v_{\alpha}(x) = e_{\alpha\beta}\partial_{\beta}\psi\,,\ \ \ \ \ \
  \omega(x) = e_{\alpha\beta}\partial_{\alpha}v_{\beta} =
  \Delta\psi\,.
\ee
They satisfy the Navier-Stokes equations
\be\label{eq:navier}
  \dot\omega + e_{\alpha\beta}\partial_{\alpha}\psi\partial_{\beta}
  \Delta\psi = \nu\Delta\omega\,,
\ee
where $\nu$ is the viscosity and $\dot\omega = \partial\omega/\partial t$.
In the case that a stirring force is present, it
would appear on the right hand side of equation (\ref{eq:navier}). As usual,
it is assumed that for large Reynolds numbers there exists an inertial range
of scales where both viscosity and stirring force can be neglected, i.e.\
all distances are much smaller than the scale of an external pump $L$, and 
much larger than the viscous scale $a$. In this case, one has the inviscid 
Hopf equation
\be\label{eq:hopf}
  \ba{c}
    {\displaystyle\sum_{p=1}^n\left\langle\omega(x_1)\ldots\dot\omega(x_p)
    \ldots\omega(x_n)\right\rangle = 0\,,}\\[1em]
    {\displaystyle\dot\omega(x) = -e_{\alpha\beta}\partial_{\alpha}\psi(x)
    \partial_{\beta}\Delta\psi(x)\,.}
  \ea
\ee
The ultraviolet cut-off $1/a$ in momentum space prohibits to go on without
a careful point splitting of the non-linear term $\dot\omega$,
\be
  \dot\omega(x) = -\overline{\lim_{a\rightarrow 0}}\ 
  e_{\alpha\beta}\partial_{\alpha}\psi(x+a)\partial_{\beta}\Delta\psi(x)\,,
\ee
where the limiting procedure includes an averaging over directions before
$|a|$ is taken to be smaller as all other distances. The key of using
CFT techniques is now to assume that $\psi$ is a primary operator of
some CFT, and to replace point split expressions by the appropriate
operator product expansion (OPE). The structure of the CFT dictates then
via the fusion rules \cite{BPZ83} which primary fields may appear on the
right hand side of the OPE. If
\be\label{eq:ope}
  \psi(x+a)\psi(x) = |a|^{2(h(\phi)-2h(\psi))}(\phi(x) + {\rm descendants})\,,
\ee
or in more compact notation of fusion rules
\be
  {}[\psi]\times[\psi] = {}[\phi] + \ldots\,,
\ee
where $\phi$ is the field with minimal conformal dimension $h(\phi)$ showing 
up in the OPE, one has
\be\label{eq:LLphi}
  \dot\omega(x) \propto |a|^{2(h(\phi)-2h(\psi))}\left[
  L_{-2}\bar L_{-1}^2 - \bar L_{-2}L_{-1}^2\right]\phi(x)\,.
\ee
Since turbulence is characterized by flux states from which conserved
quantities leak away, the CFT in question must be non-unitary, and therefore
$\phi$ is non-trivial, since only Gibbs states are described by unitary
theories.
  \par
I will now briefly summarize, what conditions on the parameters determining
the CFT are known \cite{Pol92}. Firstly, the energy density in wave number
space can be expressed by $|\psi|^2$ which yields
\be\label{eq:energy}
  E(k) \propto k^{4h(\psi)+1}\,,
\ee
and is expected to be somewhere in the range of the predictions as of Moffat 
\cite{Mof86}, $E(k)\propto k^{-11/3}$, Kraichnan \cite{Kra67},
$E(k)\propto k^{-3}$, or Saffman \cite{Saf71}, $E(k)\propto k^{-4}$.
There exist an infinite number of (within the inertial range) conserved
integrals of the type
\be
  H_n = \int\omega^n(x){\rm d}^2x\,.
\ee
The first non-trivial, $H_2$, is called enstrophy. Polyakov \cite{Pol92}
obtained from the requirement for a steady turbulence spectrum that the 
enstrophy flux be conserved (i.e.\ $\langle\dot\omega(z)\omega(0)\rangle
\propto |z|^{-h(\omega)-h(\dot\omega)}\langle\Bid\rangle$) the condition
\be\label{eq:eq}
  (h(\phi)+2) + (h(\psi)+1) = 0\,.
\ee
The issue of higher-order integrals and their fluxes has been raised
first by \cite{FaHa92,FaLe94}. 
Requiring the $H_n$ dissipation to remain constant while the
viscosity $\nu\propto a^{2h(\psi)-2h(\phi)}$ tends to zero, one obtains 
\cite{FaHa92} the condition
\be
  \dot H_n = \nu\int^{1/a}k^2h_n(k){\rm d}k \propto 
  \nu^{\frac{(n-1)(h(\psi)+1)+h(\phi)+2}{h(\phi)-h(\psi)}}\,.
\ee
Obviously, this can only be satisfied for all $n$ if $2h(\psi)=h(\phi)=-2$ 
which gives Kraichnan's spectrum. This corresponds to $\psi(x)\propto
|x|^2$ so that the vorticity $\omega$ has zero scaling dimension, maybe
logarithmic. Then, all powers of the vorticity can have constant fluxes
in $k$-space simultaneously. Therefore, Kraichnan's spectrum would satisfy
all conservation laws if it were local.
  \par
On the other hand, Polyakov \cite{Pol92} introduced an additional constraint
to ensure that the set of conformal correlators is a steady solution,
namely that the OPE (\ref{eq:ope}) vanishes in the ultraviolet limit
$a\rightarrow 0$, i.e.
\be\label{eq:eq2} 
  h(\phi) > 2h(\psi)\,,
\ee 
and thus with (\ref{eq:eq})
$h(\psi) < -1$. Note that Kraichnan's spectrum is at the limit point of the
allowed values of the conformal scaling dimensions of $\psi,\phi$. 
In \cite{FaHa92}, Falkovich and Hanany showed that this limit solution
cannot be obtained by a minimal model. Unfortunately, this limit solution
also produces a logarithmic infrared divergence after substitution into
the equations for the correlation functions.
  \par
To summarize, any candidate CFT solution carrying constant flux should
satisfy two conditions: i) the solution should be local in $k$-space, i.e.\
the integral determining the flux should converge, and ii) this constant
flux from the converging integral should be non-zero and of correct sign
to satisfy the $k$-space boundary conditions of pumping and damping.
But there is the problem, because any CFT solution satisfying (\ref{eq:eq})
and (\ref{eq:eq2}) does violate both of these conditions! To be more
precise, (\ref{eq:eq2}) tells that the energy spectrum (\ref{eq:energy})
is steeper than Kraichnan's solution, yielding a power-law infrared
divergence. Next, the second condition of non-zero flux is violated,
because the three-point function $\langle\psi\psi\psi\rangle \sim
\langle\psi\phi\rangle$ is zero since primary fields with different
dimensions $h(\psi)\neq h(\phi)$ are always orthogonal\footnote[1]{Solutions
with $\psi=\phi$ or at least $h(\psi)=h(\phi)$ were shown in \cite{FaHa92}
not to work.}.
  \par
So far, all considerations have been done under the assumption that the
vacuum expectation values of the fields (i.e.\ the one-point functions)
are all zero. If this is not the case, then there are straight-forward
modifications, e.g.\ $\langle\dot\omega(z)\omega(0)\rangle
\propto |z|^{h(\chi)-h(\omega)-h(\dot\omega)}\langle\chi(0)\rangle$, where
$\chi$ is the minimal-dimension operator in the fusion product
$[\psi]\times[\psi]\times[\psi]$. Thus, (\ref{eq:eq}) had to be replaced
by $(h(\phi)+2)+(h(\psi)+1)-h(\chi)=0$.
%
%
  \par
  \mysection{Opening the Curtain for a new Solution}
  \pn
After paraphrasing what the problems of conformal turbulence are, I come
now to the point to introduce my candidate solution. Neither is it a minimal
model, nor is it a CFT in the usual sense -- it is one of the recently
established rational logarithmic CFTs \cite{Flo95,GaKa96}, i.e.\ a CFT where 
correlation functions might be logarithmic in behavior. The fact that a CFT
can be consistently defined even in the case of logarithmic correlators was
first pointed out by Gurarie \cite{Gur93}. Further progress in understanding
such logarithmic CFTs was provided by several people, see e.g.\
\cite{CKT95,Flo95,GaKa96,Kau95,MaSe96,Roh96,ShRa96}. 
  \par
I propose that
two-dimensional turbulence can be described by the rational logarithmic
CFT with central charge $c = c_{6,1} = -24$ and maximally extended
symmetry algebra $\w(2,11,11,11)$. Actually, I will show that the
tensor product of two of these theories (having then total central charge
$c=-48$) is able to overcome most of the difficulties of the last section.
Moreover, the logarithmic behavior of certain correlation functions, and
the fact that logarithmic CFTs can posses {\em different\/} fields with equal
conformal dimensions will be quite helpful in the following.
  \par
First, I will provide the reader with the ingredients of this theory needed
below. The field content is as follows: there are 17 fields, splitting
into 12 ordinary primary fields and 5 ``logarithmic'' operators, whose $\w$ 
families we denote by $[h]$ and $[\tilde h]$ respectively, $h$ being the 
conformal dimension of the field. According to \cite{Flo95} the field
content of a $c_{p,1}$ model is given by the conformal grid of the ``would-be 
minimal model'' ${\cal M}(3p,3)$, i.e.\ the conformal weights can be expressed 
by $h_{1,r} = \frac{1}{4p}\left[(p-r)^2 - (p-1)^2\right]$ yielding the set
\be\label{eq:hwerte}{\textstyle
  \left\{[0],[-\frac{3}{8}],[-\frac{2}{3}],[-\frac{7}{8}],[-1],
  [-\frac{25}{24}],[-\tilde{1}],[-\tilde{\frac{7}{8}}],
  [-\tilde{\frac{2}{3}}],[-\tilde{\frac{3}{8}}],[\tilde{0}],[\frac{11}{24}],
  [1],[\frac{13}{8}],[\frac{7}{3}],[\frac{25}{8}],[4]\right\}}\,.
\ee
Notice that the conformal dimensions of different representations often 
are identical or differ just by integers, a characterizing feature of
logarithmic CFTs. Actually, all except the representations 
$[h_{1,p}]$ and $[h_{1,2p}]$ fall into triplets of the form
$([h_{1,r}],[\tilde h_{1,2p-r}],[h_{1,2p+r}])$, where $h_{1,2p-r}=h_{1,r}$
and $h_{1,2p+r}=h_{1,r}+r$.
The reader should consult \cite{Flo95,GaKa96} for details on the special 
structure of the indecomposable $[\tilde h]$ representations. Any of these 
splits into two 
not completely reducible representations $[\tilde h^{\pm}]$ such that the 
quantum dimensions of the both subrepresentations add up to zero. Moreover,
one has $[h_{1,r}]\subset [\tilde h_{1,2p-r}^+]$ and $[h_{1,2p+r}]\subset
[\tilde h_{1,2p-r}^-]$ with the additional relation $2[h_{1,r}]+2[h_{1,2p+r}]
=[\tilde h_{1,2p-r}]$.
  \par
The fusion rules can be calculated according to \cite{Flo95,GaKa96}.
Here, I only give the cases, which are important for the following, 
obtained by the method of \cite{Flo95}. Interpreting the right hand
sides one should keep in mind the relations between representations
mentioned above.
\be\ba{rclcl}
{}        [0]&\times&[0]         &=& [0]                                \,,\\
{}        [0]&\times&[\tilde 0]  &=& [\tilde 0]                         \,,\\
{}        [0]&\times&[-1]        &=& [-1]                               \,,\\
{}        [0]&\times&[-\tilde 1] &=& [-\tilde 1]                        \,,\\
{} [\tilde 0]&\times&[\tilde 0]  &=& 2[\tilde 0] + 2[-\frac{2}{3}] + 2[-1] +
{}                                   2[1] + 2[\frac{7}{3}] + 2[4]       \,,\\
{} [\tilde 0]&\times&[-1]        &=& 2[\tilde 0] + 2[-\frac{2}{3}] +
{}                                   [-\tilde 1] + 2[1] + 2[\frac{7}{3}]\,,\\
{} [\tilde 0]&\times&[-\tilde 1] &=& 2[\tilde 0] + 2[-\frac{2}{3}] + 2[-1] +
{}                                   2[1] + 2[\frac{7}{3}] + 2[4]       \,,\\
{}       [-1]&\times&[-1]        &=& [0] + 2[-\frac{2}{3}] + 2[-1] + 
{}                                   2[\frac{7}{3}] + 2[4]              \,,\\
{}       [-1]&\times&[-\tilde 1] &=& 2[-\frac{2}{3}] + 2[-1] + [\tilde 0] +
{}                                   2[\frac{7}{3}] + 2[4]              \,,\\
{}[-\tilde 1]&\times&[-\tilde 1] &=& 2[\tilde 0] + 2[-\frac{2}{3}] + 2[-1] +
{}                                   2[1] + 2[\frac{7}{3}] + 2[4]       \,,
\ea\ee
The only thing to learn from the fusion rules which is important is that
the leading most negative contribution in all non-trivial fusion products
is given by the representations $[-1]$ and $[-\tilde 1]$. 
  \par
The next step is to consider the tensor product of two copies of my
rational logarithmic CFT candidate -- not to be confused with the fact that
any CFT is a tensor product of a left and right chiral part. As a shorthand,
let me introduce the compact notations $\Phi_{(h|h')}(z) =
(\Phi_{h}\otimes\Phi_{h'})(z)$ for fields and $[h|h'] = [h]\otimes [h']$
for conformal families. Of course, $h(\Phi_{(h|h')}) = h+h'$. I propose now 
the (symmetrized) choices $\psi_1 = \frac{1}{2}(\Phi_{(-1|\tilde 0)} + 
\Phi_{(\tilde 0|-1)})$, $\psi_2 = \frac{1}{2}(\Phi_{(-\tilde 1|\tilde 0)} + 
\Phi_{(\tilde 0|-\tilde 1)})$ as well as $\phi_1 = \Phi_{(-1|-1)}$, $\phi_2 = 
\frac{1}{2}(\Phi_{(-1|-\tilde 1)} + \Phi_{(-\tilde 1|-1)})$, and finally 
$\phi_3 = \Phi_{(-\tilde 1|-\tilde 1)}$. The fields $\psi$ and $\phi$ which
are supposed to describe the turbulence are then linear combinations of the
fields $\psi_i$ and $\phi_j$, respectively. 
  \par
The existence of several fields
with the correct conformal dimensions $-1,-2$ and fusion products
\be
  \psi_i\times\psi_j = A_{ij}\phi_1 + B_{ij}\phi_2 + C_{ij}\phi_3 + \ldots
\ee
introduces an additional degree of freedom which might be used to cancel
certain divergences in the ultraviolet caused by logarithms. In fact,
correlation functions containing ``logarithmic'' fields (from the 
$[\tilde h]$ representations) differ from the correlation functions of 
``ordinary'' fields (from the $[h]$ representations) by multiplicative 
logarithms. The two-point functions of a pair of fields $\Phi_h$ and
$\tilde{\Phi}_h$ with $h=h_{1,r}=h_{1,2p-r}$ are
\bea\label{eq:2pointXX}
  \langle\Phi_h(z)\Phi_h(w)\rangle&=&0\,,\\
  \langle\Phi_h(z)\tilde{\Phi}_h(w)\rangle&=&\frac{C_{h\tilde h}}{(z-w)^{2h}}
    \,,\label{eq:2pointXY}\\
  \langle\tilde{\Phi}_h(z)\tilde{\Phi}_h(w)\rangle&=&\frac{1}{(z-w)^{2h}}
    \left(-2C_{h\tilde h}\log(z-w) + D_{\tilde h\tilde h}\right) 
    \,,\label{eq:2pointYY}
\eea
where the constant $D_{\tilde h\tilde h}$ is arbitrary due to the possible 
shift $\tilde{\Phi}_{h} \mapsto \tilde{\Phi}_{h} + \lambda\Phi_h$. 
If in the ultraviolet region distances become very small,
expressions of the form $(z-w)^{-\alpha}\log(z-w)\sim\frac{1}{h}
(z-w)^{-\alpha}$ asymptotically for $\alpha<0$ and $|z-w|\gg 1$. Therefore, 
choosing $D_{\tilde h\tilde h} = -C_{h\tilde h}/h$ in the definition, one can 
eliminate an ultraviolet divergence from the two-point function 
$\langle\tilde{\Phi}_h(z)\tilde{\Phi}_h(w)\rangle$. Hence, the appearance
of logarithmic operators does not spoil the nice ultraviolet behavior of 
non-unitary CFTs. The only remaining case which might cause difficulties is
$\alpha=0$ which I discuss below.
  \par
The situation in the infrared is more complicated. As with ordinary 
non-unitary CFTs, one has the problem of unphysically diverging correlation
functions. However, my candidate solution has the very special property
that the fields have {\em integer\/} dimensions. This allows non-diagonal
combinations of left and right chiral part, and in particular non-trivial
chiral states, by preserving the locality condition $h-\bar h\in\BZ$ for
a field $\Phi_h(z)\otimes\bar{\Phi}_{\bar h}(\bar z)$. These chiral
fields arise by twisting and are related to non-trivial boundary conditions.
In general, their one-point functions do not vanish. Hence, the physical 
field $\psi$ has a decomposition
\bea
  \psi(z,\bar z) &=&
  a_1\Phi_{(-1|0)}(z)\otimes\bar{\Phi}_{(-1|0)}(\bar z) +
  a_1\Phi_{(0|-1)}(z)\otimes\bar{\Phi}_{(0|-1)}(\bar z)\nonumber\\ & & +\,
  \tilde a_1\Phi_{(-1|\tilde 0)}(z)\otimes\bar{\Phi}_{(-1|\tilde 0)}(\bar z) + 
  \tilde a_1\Phi_{(\tilde 0|-1)}(z)\otimes\bar{\Phi}_{(\tilde 0|-1)}(\bar z) 
  \nonumber\\ 
  & & +\, a_{+2}\Phi_{(-1|-1)}(z)\otimes\bar{\Phi}_{(0|0)}(\bar z) +
  a_{-2}\Phi_{(0|0)}(z)\otimes\bar{\Phi}_{(-1|-1)}(\bar z)\nonumber\\ & & +\, 
  \tilde a_{+2}\Phi_{(-1|-1)}(z)\otimes\bar{\Phi}_{(\tilde 0|\tilde 0)}(\bar z)
  + \tilde a_{-2}
    \Phi_{(\tilde 0|\tilde 0)}(z)\otimes\bar{\Phi}_{(-1|-1)}(\bar z) + \ldots
\eea
and analogous for $\phi$. The same holds for other primary fields of higher
conformal weights which appear as higher order terms in the OPE. Notice that
chiral fields yield powers in $z^2$ or $\bar z^2$ instead of $|z|^4$ where the
square is due to the twofold tensor product structure I use. Thus, chiral 
fields model the additional terms in the infrared, analytical in $z^2$ or 
$\bar z^2$, which stem from the stirring forces. This is quite natural, since 
stirring forces which generate vortices and turbulence are chiral. Taking all 
this into account one gets
\bea
  \lefteqn{\langle\psi(z)\psi(0)\rangle}\nonumber\\ 
  &=& |z|^{-4h(\psi)}\langle\Bid\rangle +
    c_1|z|^{2(h(\phi)-2h(\psi))}\langle\phi\rangle + 
    c_{+2}z^{h(\chi)-4h(\psi)}\langle\chi^+\rangle +
    c_{-2}\bar z^{h(\chi)-4h(\psi)}\langle\chi^-\rangle + \ldots\nonumber\\
  &=& |z|^{-4h(\psi)}\left[ 1 + c_1\left(\frac{|z|}{L}\right)^{2h(\phi)}
    \langle\phi\rangle + 
    c_{+2}\left(\frac{z}{L}\right)^{h(\chi)}\langle\chi^+\rangle +
    c_{-2}\left(\frac{\bar z}{L}\right)^{h(\chi)}\langle\chi^-\rangle +
    \ldots \right]\,,
\eea
where the last equation reintroduces the infrared scale. Since $h(\psi)=-1$ 
and $h(\phi)=-2$, the second term is independent of $|z|$. The chiral fields 
$\chi^{\pm}$ stand for the next leading orders and have $h(\chi)\in 2\BZ$,
$h(\chi)\geq -2$.
Therefore, after a Fourier transform, one arrives at
\be
  \langle\psi(k)\psi(-k)\rangle =
    \frac{\alpha_0}{|k|^{2-4h(\psi)}} + 
    \frac{\alpha_1}{L^{2h(\phi)}}\delta(k) + 
    \frac{\alpha_2}{L^{h(\chi)}}\frac{\partial^{h(\chi)-4h(\psi)}}
      {\partial k^{h(\chi)-4h(\psi)}}\delta(k) + \ldots
\ee
In my case of $h(\psi)=-1$, $h(\phi)=-2$, this precisely resembles
Polyakov's Bose condensate in momentum space, formed by the large
scale motions. The main difference to Polyakov's approach is that
the Bose condensate is contained {\em within\/} the CFT description.
The first non-trivial higher order term comes from the {\em chiral\/} primary 
fields of the $[-1|-1]$ and $[-\tilde 1|-\tilde 1]$ representations. 
The coefficients can arbitrarily be adjusted by 
rescaling fields. Notice that all fields $\chi$ are given as linear 
combinations of different possible decompositions such that one can always
enforce $\langle\chi\rangle = 0$ if so required by the boundary conditions.
  \par
This might help to solve the infrared problem, but it seems that now the
ultraviolet region spoils the solution, since (\ref{eq:ope}) and thus
(\ref{eq:LLphi}) do not vanish in the limit $a\rightarrow 0$. Luckily,
there is an interesting way out. Namely, if the pair of fields
$\Phi_h$, $\tilde{\Phi}_h$ has integer valued conformal weights, $h\in\BZ$,
then the logarithmic CFT furnishes a hidden continuous symmetry \cite{CKT95},
and $\Phi_h$ can be regarded as a conserved current due to 
(\ref{eq:2pointXX}--\ref{eq:2pointYY}).
As has been pointed out by Polyakov himself \cite{Pol92}, one way to 
satisfy the inviscid Hopf equations (\ref{eq:hopf}) is that the operator
\be
  \Omega(z) = L_0\bar L_0
    \left[L_{-2}\bar L_{-1}^2 - \bar L_{-2}L_{-1}^2\right]\phi(z)
\ee
is a symmetry of the underlying CFT. Such a symmetry 
precisely appears in my example, since $\phi$ is composed out of
$\Phi_{h_{1,5}}$ and $\tilde{\Phi}_{h_{1,7}}$ which 
form such a pair with conformal weights $h_{1,5}=h_{1,7}=-1$, where 
$\Phi_{h_{5,1}}$ gives rise to the hidden continuous symmetry.
The additional $L_0\bar L_0$ term has been introduced to ensure that
the Hopf equations only feel $\Phi_{h_{1,5}}$ in the correlators such
that the symmetry argument applies. To see this, one notes that \cite{ShRa96}
\be 
  L_0\langle\tilde{\Phi}_1(z_1)\tilde{\Phi}_2(z_2)\Phi_3(z_3)\ldots
    \Phi_n(z_n)\rangle = k_n\langle\Phi_1(z_1)\Phi_2(z_2)\Phi_3(z_3)\ldots
    \Phi_n(z_n)\rangle\,,
\ee
where $\Phi_i$ denote ordinary primary fields, $\tilde{\Phi}_i$ logarithmic
operators, and $k_n$ are universal constants. In my 
point of view it is appealing to find a possible solution of the inviscid Hopf 
equations, which is not trivial, i.e.\ provided by a symmetry rather than the 
fact $\Omega\equiv 0$. 
%
%
  \par
  \mysection{Happy End?}
  \pn
Let me summarize what has been accomplished so far. The proposed CFT solution
is a rational logarithmic CFT with $c=-48$. It furnishes Kraichnan's picture
of two-dimensional turbulence and incorporates automatically the stirring
forces in the infrared region via chiral primary fields. The Hopf equations 
are satisfied due to an exact symmetry of this CFT, and are therefore valid 
both, in the ultraviolet as well as in the infrared. 
  \par
However, I do not expect that this is the whole story. Polyakov \cite{Pol92}
argued that two-dimensional turbulence is conformally invariant. But in an
incompressible fluid it should also be invariant under area-preserving
diffeomorphisms, generated by the $\w_{1+\infty}$ algebra. Therefore, I
conjecture that my logarithmic CFT can be extended to an non-unitary 
$\w_{1+\infty}$ model. It has been shown in \cite{ShRa96} that at least
the Borel subalgebra of $\w_{1+\infty}$ is a symmetry of logarithmic CFTs.
This might link two-dimensional turbulence with the transitions between
fractional quantum Hall states. For first ideas in this direction see
\cite{Flo96}.
  \par
It has been pointed out in \cite{FaHa92} that ordinary rational CFTs, in
particular minimal models, can never yield Kraichnan's turbulence, although
the latter can be approximated by certain sequences of minimal models.
Their sequence of minimal models which approaches $h(\phi)=-2$ has a limit
of central charge $c=-47$ under the assumptions that $\phi$ is the field
of (almost) lowest conformal weight in the operator algebra. In general
one obtains
\be\label{eq:pq}
  \left(\frac{p}{q}\right)_{\pm} =
  (1-2h) \pm \sqrt{4(h^2-2) + \frac{n^2}{q^2}}\,,
\ee
as the condition for a minimal model with $c_{p,q}$ to contain a field of
conformal dimension $h=h(p,q)_{r,s}$ such that $(ps-qr)^2 = n^2$. There are,
however, two ways to make the right hand side of (\ref{eq:pq}) independent
of $q$ for $q\gg 1$. Either $n$ is kept always small compared to $q$,
or $n = aq+b$ with $a,b$ small compared to $q$. The first way yields
$p/q = (1-2h) \pm 2\sqrt{h^2-h}$, which in general is irrational. This means
that the limiting point of such a sequence of minimal models is not a
rational CFT (although, there exist rational CFTs at the limiting
central charges $c=1+24h$, see \cite{Flo92}). On the other hand, the second 
way might yield for instance
$p/q = 2(1-2h)$, thus $p,q$ become non coprime in this limit. 
  \par
As I have shown
in my earlier work \cite{Flo95}, this indicates that the limiting point
of such a sequence of minimal models is a rational logarithmic CFT. One 
easily can check that there are such $c_{p,1}$ limiting points for
$h=-1$ and $h=-2$. However, there is no $c_{p,1}$ model which allows both
conformal weights at the same time. Hence, I was forced to consider
tensor products. Considering $\psi\times\psi=\phi$ it is a simple task to
convince oneself that the twofold tensor product of the $c_{6,1}$ model is 
the only possibility.
  \par
One of the appealing properties of ``Kraichnan's spectrum''\footnote[1]{to 
be precise, the non-trivial generalization of it given in \cite{FaLe94}} is 
that it respects all conservation laws given by the $H_n$ integrals of 
motion. If one would insist on a different spectrum such as 
$E(k)\propto k^{-4}$, one will not find any suitable candidate within the
$c_{p,1}$ models: Using a tensor product construction one is bound to
the condition $h(\phi)=2h(\psi)$ and to Kraichnan's spectrum. In general, a
spectrum $E(k)\propto k^{-\alpha}$ means $h(\psi) = -(\alpha+1)/4$. 
To find such a field in a $c_{p,1}$ model, one needs $h(p,1)_{1,s}$ where
\be
  s = p \pm \sqrt{p^2-(3+\alpha)p+1}
\ee
must be an integer $1\leq s\leq 3p-1$. If one hopes to find also the field
$\phi$ in the same theory, one had to satisfy the additional Diophantine
equation
\be
  s' = p \pm \sqrt{p^2-(13-\alpha)p+1}\,.
\ee
If the case $h(\psi)=h(\phi)=-\frac{3}{2}$ is excluded due to the unphysical
spectrum $E(k)\propto k^{-5}$, there are no further simultaneous solutions
of both Diophantine equations. From this it is concluded that, if my solution 
is correct, then it is unique.
  \par
I mentioned two problems in the first section, which usually are not solved
by a CFT approach to two-dimensional turbulence. 
  \par
Let me start with the problem 
of non-zero flux. In my theory, the three-point function 
$\langle\psi\psi\psi\rangle \sim \langle\psi\phi\rangle$ does not necessarily 
vanish, because both, $\phi$ and $\psi$, are made out of the same basic 
fields, namely $\Phi_{-1}$, $\tilde{\Phi}_{-1}$, $\Phi_0$ and $\tilde{\Phi}_0$.
Therefore, the three-point function reduces to terms (notice the factorization
due to the tensor product)
\be
  \langle\psi\psi\psi\rangle\sim
  \langle\Phi_{-1}\Phi_{-1}\rangle\langle\Phi_{-1}\Phi_0\rangle + 
  \langle\Phi_{-1}\Phi_{-1}\rangle\langle\Phi_{-1}\tilde{\Phi}_0\rangle + 
  \ldots
\ee
There are several non-vanishing contributions. If the one-point functions 
do not vanish then e.g.\ the first term would contribute. In any case,
the second term does not vanish due to the properties of logarithmic
operators \cite{Flo95,GaKa96,MaSe96,ShRa96}.
  \par
The other problem is the locality of the spectrum, i.e.\ the convergence
of the integral determining the flux in $k$-space. As derived e.g.\ in 
\cite{FaHa92}, one has to consider the integral
\be
  \int_{1/L} kh_2(k){\rm d}k\,,
\ee
where the density $h_2(k) \propto k^{2h(\psi)}$. Kraichnan's spectrum yields
a logarithmic divergence for the infrared scale $L\rightarrow\infty$. If one
keeps in mind that the theory in question has logarithmic operators, one has
the modification $h_2(k) \propto k^{2h\psi)}(C + C'\log(k/a))$. Notice that a
logarithm is always scale dependent. If the infrared and the ultraviolet
scales are related such that their product $La$ remains constant, then
the integral converges for $C=0$ in the infrared region. It seems quite 
natural to use the scales set by physics to define the scale of the 
logarithmic terms.
  \par
One additional feature of my solution is that magneto-hydrodynamics in
two dimensions is contained in my theory as a mere corollary. Since all
fields in question are degenerated such that there are inequivalent fields
of equal conformal weights, the theory has enough ``space'' to
carry the magnetic degrees of freedom. Besides the stream function $\psi$
there is a second independent dynamical variable, the magnetic flux function
$\tilde{\psi}$. Due to the Alf'ven effect,
kinetic energy and magnetic energy are asymptotically exact equi-partitioned
within the inertial range \cite{Iro63,Kra65}. Therefore, the
corresponding fields must have the same conformal weights, i.e.\
$h(\psi)=h(\tilde{\psi})$. It already has been noted in \cite{RaRo95} that 
the Alf'ven effect brings logarithmic correlators onto the stage, naturally 
accounted for in my theory. Details are left for future work.
 \par
To conclude, I would like to mention one possible serious drawback of the CFT 
approach to two-dimensional turbulence: there is no time. This means that
a CFT approach is blind for any renormalization due to
spectral transfer of vorticity flux to smaller scales, driven by the strain
of velocity shear. Fortunately, it turns out \cite{FaLe94} that the
renormalization law, determined from time correlations between velocity
gradients from different spectral intervals, is just provided by the mean
stretching rate (Lyapunov exponent) $\bar{\lambda}$. In my approach
$\xi=\log(|z|/L)$ is the fundamental variable, since
$\omega(z)$ has scaling dimension zero. Notice that any logarithmic CFT can
only yield integer powers of $\log(z)$ terms. The 
renormalization is then simply given by $\xi\mapsto\xi/\bar{\lambda}(\xi)$,
where one has as a result \cite{FaLe94} that $\bar{\lambda}(\xi) \propto 
\xi^{1/3}$. With this replacement of the fundamental variable $\xi$, my 
approach reproduces the results for $n$-point correlators, e.g.\ 
$\langle\omega_1\omega_2\ldots\omega_{2n}\rangle \propto \xi^{2n/3}$, as
given in \cite{FaLe94}. The energy spectrum renormalizes then to 
$E(k)\propto k^{-3}\log^{-1/3}(kL)$ which was derived by Kraichnan 
\cite{Kra67} assuming weak time correlations and using a one-loop 
approximation. 
  \par
I have now to leave it to the estimated reader whether they lived happily
ever after $\ldots$
  \bigskip\pn
{\bf Acknowledgment:} I would like to thank  
V.~Gurarie, A.~Polyakov and in particular G.~Falkovich and A.~Hanany 
for discussions and comments.
This work has been supported by the Deutsche Forschungsgemeinschaft.
%
%
  
  \end{document}